\documentclass[preprint2,twocolappendix]{aastex631}
\usepackage[version=4]{mhchem}
\usepackage{physics}
\usepackage{comment}
\usepackage{CJK}
\newcommand{\meyr}{M$_{\oplus}\cdot\rm{yr}^{-1}$}

\newcommand{\mj}{M$_{J}$}
\newcommand{\rj}{R$_{J}$}
\newcommand{\E}{E_{r}}
\newcommand{\F}{F_{r}}
\newcommand{\kp}{\kappa_{\rm{P}}}
\newcommand{\kr}{\kappa_{\rm{R}}}
\newcommand{\ar}{a_{\rm{R}}}
\newcommand{\tg}{T_{\rm{g}}}
\newcommand{\trad}{T_{\rm{rad}}}
\newcommand{\gcmc}{g$\cdot$cm$^{-3}$}

\shorttitle{Planetary Accretion Shocks with a Realistic EoS}
\shortauthors{Chen \& Bai}
\graphicspath{{./}{figures/}}
\begin{document}
\begin{CJK*}{UTF8}{gbsn}

%\title{Accretion shocks of forming gas giants}
\title{Planetary Accretion Shocks with a Realistic Equation of State}

\correspondingauthor{Zhuo Chen}
\email{zc10@ualberta.ca}
\correspondingauthor{Xuening Bai}
\email{xbai@tsinghua.edu.cn}

\author[0000-0001-7420-9606]{Zhuo Chen (陈卓)}
\affiliation{Department of Astronomy, Tsinghua University \\
Beijing 100084, China}

\author[0000-0001-6906-9549]{Xuening Bai (白雪宁)}
\affiliation{Department of Astronomy, Tsinghua University \\
Beijing 100084, China}
\affiliation{Institute for Advanced Study, Tsinghua University \\
Beijing 100084, China}

\begin{abstract}

The final stage of gas giant formation involves accreting gas from the parent protoplanetary disk. In general, the infalling gas likely approaches a free-fall velocity, creating an accretion shock, leading to strong shock heating and radiation. We investigate the kinematics and energetics of such accretion shocks using 1D radiation hydrodynamic simulations. Our simulations feature the first self-consistent treatment of hydrogen dissociation and ionization, radiation transport, and realistic grey opacity. By exploring a broad range of giant planet masses (0.1-3\mj) and accretion rates ($10^{-3}$-$10^{-2}$\meyr), we focus on global shock efficiency and the final entropy of the accreted gas. We find that radiation from the accretion shock can fully disassociate the molecular hydrogen of the incoming gas when the shock luminosity is above a critical luminosity. 
Meanwhile, the post-shock entropy generally fall into ``cold" ($\lesssim12k_{\rm{B}}/m_{\ce{H}}$) and ``hot" ($\gtrsim16k_{\rm{B}}/m_{\ce{H}}$) groups which depends on the extent of the endothermic process of \ce{H2} dissociation.
While 2D or 3D simulations are needed for more realistic understandings of the accretion process, this distinction likely carries over and sheds light on the interpretation of young direct imaging planets.
\end{abstract}


\section{Introduction} \label{sec:intro}

A major paradigm shift in our understandings of protoplanetary disks (PPD) over the past few years is the ubiquity of disk substructures found in high-resolution observations of protoplanetary disks \cite[e.g.,][]{marel2013,dipierro2015,perez2016,andrews2018,long2018,avenhaus2018}. As a common interpretation, these substructures are considered as the outcome of planetary-mass companions interacting with the disk, which generally requires the formation of giant planets which can naturally open gaps (e.g., \citealp{goldreich1979,lin1986,goodman2001,dong2017}), create asymmetries \cite[e.g.,][]{valborro2007,zhu2014}, or drive spiral patterns \cite[e.g.,][]{dong2015,bae2018,bae2021}. Despite the expectation of multiple planets carving substructures in dozens of disks, including evidence from kinematic signatures in a few systems \cite[e.g.,][]{pinte2018,pinte2020,teague2019}, so far the only bona fide detection has been the two giant planets in the central cavity of the PDS 70 disk \citep{keppler2018,wagner2018,haffert2019}. The two planets are found to be accreting from the parent PPD as the pathway to build up their masses, showing H$\alpha$ emission characteristic of planetary accretion shocks \citep{aoyama2018,aoyama2019,thanathibodee2019,szulagyi2020}.

The ability to detect young giant planets, and direct imaging of gas giants in general after disk dispersal, crucially depends on its initial conditions, particularly the initial entropy that largely sets the subsequent evolution of its luminosity and temperature as main observables.
The initial entropy is closely related to the amount of the specific energy (energy per unit mass) of the accreted gas. In general, the smaller the fraction of specific energy retained in the gas, the lower the initial entropy.
Correspondingly, models of giant planet evolution are classified into ``cold-", ``warm-", and ``hot-start" models that primarily differ in their initial entropy \citep{marley2007,spiegel2012}. The differences among these models are most pronounced over the first to a few tens of Myrs after giant planet formation \citep{berardo2017}, and hence crucially affect the inference of detection limits and estimation of planet masses.

In the standard theory of giant planet formation by core accretion \citep{pollack1996}, the bulk of giant planet mass is built up by accreting gas from PPD in a runaway manner. Upon becoming sufficiently massive, the planet opens a gap and become ``detached" from the disk, forming a circumplanetary disk (CPD, e.g., \citealp{ayliffe2009,tanigawa2012,gressel2013}). How gas accretion proceeds through a CPD is unclear, but existing studies suggest that 
most gas fall into the circumplanetary region from high latitudes from the meridional flow \cite[e.g.,][]{szulagyi2014,szulagyi2017,fung2019}.
Planetary accretion may proceed through an accretion shock either from this infalling material, or via magnetospheric accretion from the CPD. This accretion shock is also the source of the H$\alpha$ emission observed from the PDS 70 system.

Detailed models of planetary accretion shocks is of great significance to understand the initial conditions of giant planets. With the accreting material approaching at near free-fall velocity, such models are usually in 1D, focusing on energy transport through the shock and the accreting column 
\citep{marleau2017,marleau2019}. In this letter, we show that a realistic equation of state (EoS), that incorporates the dissociation of \ce{H2}, plays a highly significant role in setting the post-shock entropy of the accreting gas that will further affect the initial conditions of giant planet evolutionary models.

This letter is organized as follows. Section \ref{sec:model} describes the setup and physical ingredients of our 1D radiation hydrodynamic simulations. Major results are presented in Section \ref{sec:result}, and are summarized with further discussion in Section \ref{sec:conclusion}.

\section{Physical model} \label{sec:model}

In this section, we describe the setup for our accreting gas giants simulations, highlighting the combination of radiation
hydrodynamics with a general EoS that incorporates the dissociation of hydrogen molecules.

\subsection{Governing equations}

We solve radiation hydrodynamic equations in 1D spherical geometry along the radial ($r$) direction with a general EoS, for gas accretion towards a planet with mass $M_p$. The governing equations are
\begin{eqnarray}
	\pdv{\rho}{t}+\frac{1}{r^2}\pdv{r}(r^{2}\rho v)&=&0\ ,		\\
	\pdv{\rho u}{t}+\frac{1}{r^2}\pdv{r}(r^{2}\rho v^{2})&=&-\pdv{p}{r}-\frac{\rho G M_{p}}{r^{2}}\ ,	\\
	\pdv{E}{t}+\frac{1}{r^2}\pdv{r}[r^{2}(E+p)v]&=&\mathbb{G}-\frac{\rho G M_{p}}{r^{2}}v\ ,	\label{eqn:energyhydro}\\
	\pdv{\E}{t}+\frac{1}{r^2}\pdv{r}(r^{2}\F)&=&-\mathbb{G}\ ,	\label{eqn:energyrad}
\end{eqnarray}
where $\rho$, $v$, $p$, $E$, and $G$ are gas density, radial velocity, pressure, total energy density, and gravitational constant, with
\begin{eqnarray}
	p&=&\sum_i n_i k_{\rm{B}}\tg\ ,	\\
	E&=&\rho(\epsilon_{\rm kin}+\epsilon)=\rho(v^{2}/2+\epsilon)\ ,
\end{eqnarray}
where $n_i$ is the number density of the $i$th species (to be specified later),
$k_{\rm{B}}$, $\tg$, $\epsilon_{\rm kin}$, $\epsilon(\rho,\tg)$ are Boltzmann constant, gas temperature, specific kinetic energy and specific internal energy.
Radiation energy density and energy flux are denoted by
$E_r$ and $F_r$, and $\mathbb{G}$ is the rate of energy exchange between radiation and matter, given by
\begin{eqnarray}
	\mathbb{G}&=&\kp\rho c(\E-\ar\tg^{4})\ ,
\end{eqnarray}
where $\kp$, $\ar$ and $c$ are Planck opacity, radiation constant and speed of the light.
For future convenience, radiation temperature $T_{\rm rad}$ is defined by $E_r=\ar T_{\rm rad}^4$.

The radiation sub-system is closed using the flux-limited diffusion (FLD) approximation, which relates $\E$ and $\F$ by
\begin{eqnarray}
	\F &=&-\frac{c\lambda(R)}{\kr\rho}\pdv{\E}{r}\ ,	\\
	R&=&\frac{|\partial\E/\partial r|}{\kr\rho\E}\ \label{eqn:R} ,
\end{eqnarray}
where $\kr$ is the Rosseland mean opacity. In this paper, we adopt the flux limiter $\lambda(R)$ described in
\cite{levermore1981}
\begin{equation}
	\lambda(R)=\frac{2+R}{6+3R+R^2}\ .
\end{equation}

\subsection{Equation of state and opacity}\label{sec:eos}

Our typical simulations encompass a temperature range between $\sim10^{2-4}$K, where $\ce{H2}$ can be dissociated
and eventually ionized, and standard ideal gas EoS becomes questionable. As a first study, we incorporate
such physics but make a simplified assumption of chemical local-thermal equilibrium (LTE) of \ce{H2}, \ce{H}, and \ce{H+}.
The abundance of these species can be obtained analytically from the Saha equations at runtime according to Appendix
C of \cite{chen2019}. 
The EoS is given by $\epsilon=\epsilon(\rho,\tg)$ in an analytical from, which avoids the use of a tabulated EoS and improves the efficiency and accuracy in our simulations.
In particular, we consider a hydrogen mass fraction of $X=0.74$, assuming remaining mass in helium.

One important quantity we compute in this work is the gas entropy. With the general EoS, it is given by
\begin{equation}
	s=\sum_{i}\frac{k_{\rm{B}}n_{i}}{\rho}\left(1+\frac{d\ln Z_{i}}{d\ln T}-\ln\frac{n_{i}}{Z_{i}}\right),
\end{equation}
where $Z_i$ is the partition function for the $i$th species, given in Appendix \ref{appendix:eos}.

Opacity is a crucial physical component in our model. We adopt the same opacity tables as in \cite{marleau2019}, combining the gas opacity table of \cite{malygin2014} that dominates over $1500$K, and the dust opacity table of \cite{semenov2003} that dominates below 1100-1200K, depending on the density.
For temperatures in between, the maximum of the value of the two tables is taken.

Note that the opacity tables assume gas and radiation temperatures are the same. In reality, as we will see, $\trad$ and $\tg$ can be different at the Zel'dovich spike and \ce{H2} dissociation region. Here we use the radiation temperature to obtain opacity from lookup tables, bearing in mind the caveat which can be improved in future works.

\centerwidetable
\begin{deluxetable*}{cccccc|cc|cccccccc}
\label{tab:sum}
\tablenum{1}
\tablecaption{Simulation parameters and results.}
\tablehead{
\colhead{ID}	&
\colhead{$M_{p}$}   &
\colhead{$\dot{M}_{p}$} &
\colhead{$r_{\rm{in}}$} &
\colhead{$p_{\rm{in}}$} &
\colhead{$r_{p}$} &
\colhead{$s_{\rm{ps}}^{l8}$} &
\colhead{$s_{\rm{ps}}^{l7}$} &
\colhead{$\eta^{\rm{phy}}$}		&
\colhead{$\chi_{\ce{H},\rm{pre}}$}	&
\colhead{$\chi_{\ce{H},\rm{ps}}$}	&
\colhead{$\mathcal{M}_{\rm{pre}}$} &
\colhead{$T_{g,\rm{pre}}$}	&
\colhead{$T_{g,\rm{ps}}$}   &
\colhead{$p_{\rm{ps}}$}   &
\colhead{$\frac{\dot{M}_{p}}{{\dot{M}_{\rm{crit}}}}$} \\
\colhead{}	&
\colhead{(\mj)}  &
\colhead{$\big(\frac{M_{\oplus}}{\rm{yr}}\big)$}  &
\colhead{(\rj)}   &
\colhead{(bar)} &
\colhead{(\rj)} &
\colhead{$\big(\frac{k_{\rm{B}}}{m_{\ce{H}}}\big)$} &
\colhead{$\big(\frac{k_{\rm{B}}}{m_{\ce{H}}}\big)$} &
\colhead{\%}	&
\colhead{\%}	&
\colhead{\%}	&
\colhead{}	&
\colhead{(K)}	&
\colhead{(K)}   &
\colhead{($\frac{\rm{bar}}{10^{3}}$)}   &
\colhead{}
}
\decimalcolnumbers
\startdata
1   &   0.1     &   $10^{-2}$   &   1.0     &   2   & 1.284	&	  11.42  	&	11.42	&	82.23	&	16.67	&	5.18	&	5.24	&	1865	&	1920    &   2.850    &   0.56	\\
2	&   0.3     &   $10^{-2}$   &   1.1     &   5   & 1.294	&	  17.04  	&	17.06	&	64.90	&	98.73	&	71.55	&	6.00	&	2458	&	2474    &   5.077    &   1.28	\\
3	&   1     &   $10^{-2}$   &   1.5     &   10   & 1.679	&	  19.21  	&	19.20	&	81.59	&	99.93	&	95.72	&	8.39	&	2685	&	2754    &   4.987    &   2.46	\\
4	&   1     &   $10^{-2}$   &   1.7     &   10    & 1.865	&	  18.25  	&	18.20	&	82.34	&	99.72	&	83.39	&	8.61	&	2495	&	2526    &   3.765    &   1.81	\\
5	&   3     &   $10^{-2}$   &   1.7     &   10     & 1.780	&	 19.61  	&	19.61	&	92.81	&	100		&	99.92	&	12.57	&	3358	&	3465    &   7.843    &   7.05	\\
6	&   3     &   $10^{-2}$   &   1.9     &   10     & 1.998	&	 19.68  	&	19.67	&	92.12	&	100		&	99.74	&	12.37	&	3092	&	3201    &   5.791    &   4.95     \\
7	&   0.1     &   $10^{-3}$   &   1.0     &   2   & 1.266	&	  11.45  	&	11.45	&	91.85	&	0		&	0		&	6.06	&	1290	&	1299    &   0.304    &   0.06  	\\
8	&   0.3     &   $10^{-3}$   &   1.1     &   5   & 1.198	&	  11.24  	&	11.23	&	97.02	&	1.24	&	0.11	&	11.82	&	1456	&	1468    &   0.696    &   0.24	\\
9	&   1     &   $10^{-3}$   &   1.5     &   10   & 1.568	&	 	11.49  	&	11.48	&	98.52	&	11.76	&	1.21	&	18.04	&	1612	&	1667    &   0.702    &   0.37	\\
10	&   1     &   $10^{-3}$   &   1.7     &   10   & 1.785	&	 	11.47  	&	11.48	&	98.57	&	4.43	&	0.35	&	17.83	&	1511	&	1536    &   0.489    &   0.25	\\
11	&   3     &   $10^{-3}$   &   1.7     &   10   & 1.734	&	 	13.16	&	13.21	&	98.34	&	71.71	&	20.41	&	23.12	&	1831	&	2040    &   1.130    &   0.81  	\\
12	&   3     &   $10^{-3}$   &   1.9     &   10   & 1.940	&	 	12.18	&	12.21	&	98.92	&	48.3	&	8.18	&	24.02	&	1740	&	1889    &   0.817    &   0.58
\enddata
\tablecomments{From column 1 to column 16: (1) the model ID, (2) planet mass, (3) accretion rate, (4) inner boundary radius, (5) inner boundary pressure, (6) radius where $p_{\rm{ram}}=\rho v^{2}=p$, (7) post-shock entropy with 8 SMR levels, (8) post-shock entropy with 7 SMR levels, (9) global shock efficiency, defined in Equation \ref{eqn:eta}, (10) the number fraction of \ce{H} at the pre-shock radius, (11) the number fraction of \ce{H} at the post-shock radius, (12) pre-shock gas Mach number, (13) pre-shock gas temperature, (14) post-shock gas temperature, (15) post-shock pressure, (16) and the actual accretion rate divided by the critical accretion rate.}
\end{deluxetable*}

\subsection{Simulation setup}

We solve the numerical problem with {\tt Guangqi} \citep{chen2021}, a new 1D radiation-hydrodynamic code with adaptive- and static-mesh-refinement (AMR/SMR). It employs the FLD approximation for grey radiative transfer, which is solved implicitly and is self-consistently coupled with a general EoS. In this problem, the shock and the planetary atmosphere require high resolution. They are located at the bottom of the computational domain and we employ SMR to properly resolve the Zel'dovich spike behind the shock (Section \ref{sec:sps}).

Our simulation domain spans between $[r_{\rm in}, r_{\rm max}]$, where $r_{\rm max}$ is fixed to 20\rj\ (\rj\ is Jupiter radius), using an uniform grid with 256 cells at base level. We use 7 or 8 levels of mesh refinement with each level doubling the resolution of the parent level. Therefore, the finest cell has a length of approximately 43.6 km or 21.8 km ($r_{\rm{in}}$ varies from 1 to 1.9\rj). In comparison, the finest cell in \cite{marleau2019} is 35.7 km.
\begin{comment}
The atmospheric scale height can be calculated by,
\begin{eqnarray}
    H&=&\frac{k_{\rm{B}}\tg r^{2}}{M_{p}G\mu m_{\ce{H}}}    \\
        &=&216\left(\frac{M_{J}}{M_{p}}\right)\left(\frac{\tg}{1500\rm{ K}}\right)\left(\frac{r}{r_{J}}\right)^{2}\left(\frac{\mu}{2.3}\right)\rm{km},
\end{eqnarray}
where $\mu$ is the mean atomic weight of the gas. In our calculation, one scale height typically covers 10 or more cells and is sufficiently resolved.
\end{comment}

Gas is injected from the outer boundary assuming free-fall velocity, and
gas density at the outer boundary is parameterized by the accretion rate $\dot{M}_p$, given by
\begin{eqnarray}
	v_{\rm{inj}}&=&-\sqrt{\frac{2GM_{p}}{r_{\rm{max}}}}\ ,	\\
	\rho_{\rm{inj}}&=&-\frac{\dot{M}_{p}}{4\pi f_{c} r_{\rm{max}}^{2}v_{\rm{inj}}}\ ,\label{eq:rhoinj}
\end{eqnarray}
for spherical geometry, where $f_c$ is the covering fraction of the accretion flow over planetary surface, and we take $f_c=1$ in this work.
%In this work, we have assumed that accretion is spherically symmetric and $f=1$.
Gas temperature in the outer boundary is linearly extrapolated at runtime, and radiation energy is set according to $\partial (r^2\E)/\partial r=0$.

For simplicity, we set the initial condition to be $\rho_{\rm{init}}=\rho_{\rm{inj}}$, $v_{\rm{init}}=v_{\rm{inj}}$, and $T_{\rm{g,init}}=T_{\rm{rad,init}}=100$ K everywhere. The particular form of initial condition does not affect the steady state solution that we look for after running the simulations for many free-fall timescales.

Setting inner boundary conditions requires some care. A straightforward reflecting boundary condition would gradually build up mass and pressure near the inner boundary, and the system hardly achieves a steady state. In reality (especially with $f_c<1$), the post-shock gas flow likely spreads out, eventually maintaining equilibrium with atmospheric pressure. This motivates us to choose a fixed-state inner boundary condition set by $[\rho_{\rm{in}},p_{\rm{in}},T_{\rm{g,in}}]$ in gas variables. However, $\rho_{\rm{in}}$, $p_{\rm{in}}$, and $T_{\rm{g,in}}$ are unknown beforehand. In practice, we choose $p_{\rm in}$ as representative atmospheric pressure at planet surface for each simulation (to be specified in the next subsection). We first set the inner hydro boundary condition to be non-penetrating and the inner radiation boundary condition to be zero gradient. As the gas falls onto the planet, we keep monitoring the increase of gas pressure at the inner boundary. Once it reaches $p_{\rm{in}}$, we record the $\rho$ and $\tg$ of the innermost cell as $\rho_{\rm{in}}$, and $T_{\rm{g,in}}$, and impose the aforementioned fixed-state inner boundary condition. The radiation boundary condition is still zero gradient.

\begin{figure*}
	\centering
	\includegraphics[width=1.04\columnwidth]{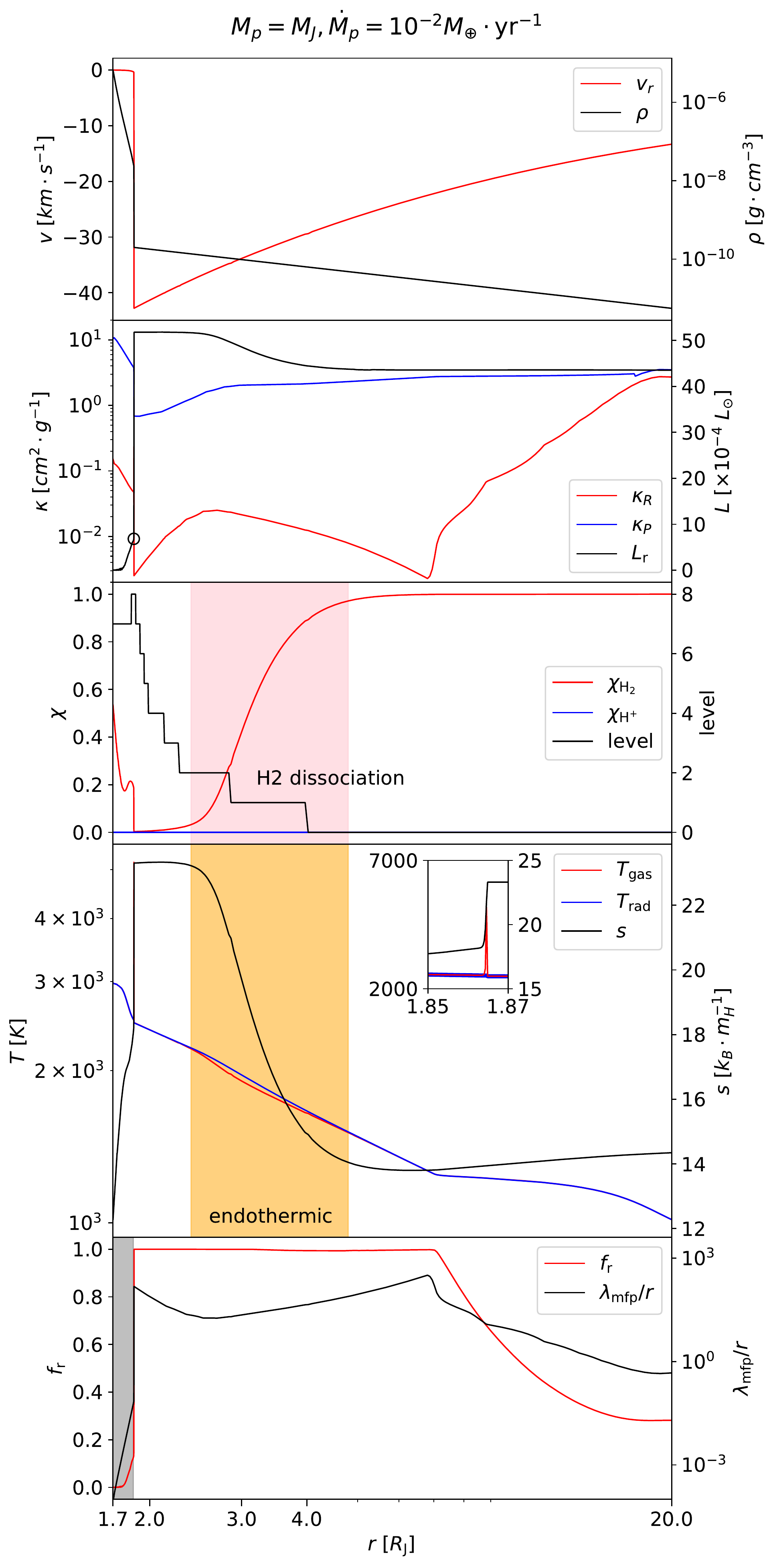}
	\includegraphics[width=1.04\columnwidth]{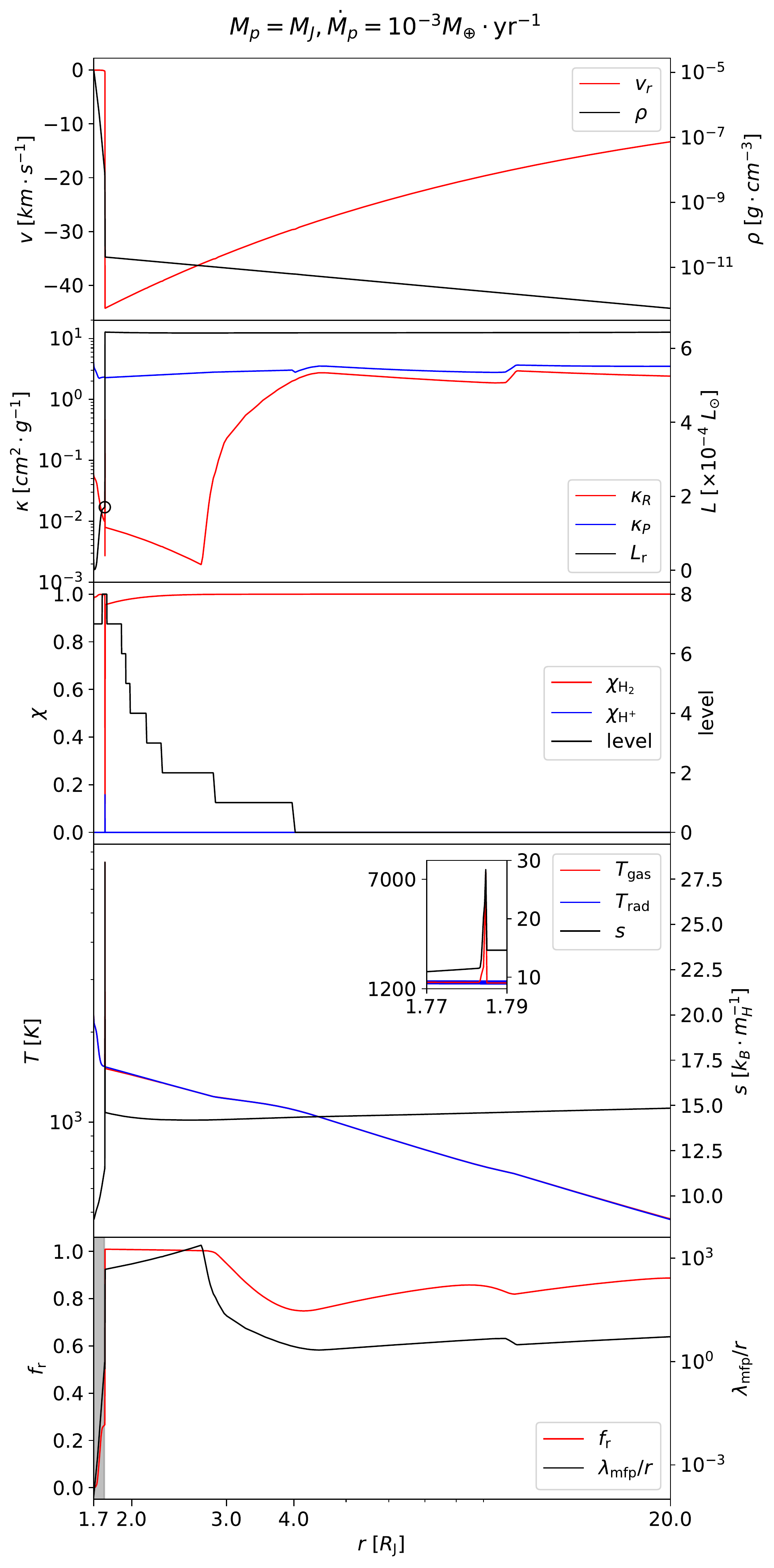}
	\caption{The shock profiles of a Jupiter mass planet accreting at high (model 4, left) and low (model 10, right). From top to bottom, the first panel: velocity and density profiles in red and black lines. Second panel: $\kr$ and $\kp$ profiles in the red and blue, and $L_{r}$ profile in the black line. The value of $L_{r}$ at the post-shock radius is marked with a black circle. Third panel: $\chi_{\ce{H2}}$ and $\chi_{\ce{H+}}$ profiles in red and blue lines. SMR levels are shown in black. The pink color covers the region of \ce{H2} dissociation. Fourth panel: $\tg$ and $\trad$ profiles in red and blue lines, with entropy profile $s$ shown in black. Hydrogen dissociation is an endothermic process and marked in orange. Fifth panel: $f_{\rm{r}}$ defined in Equation \ref{eqn:reduced} and $\lambda_{\rm{mfp}}/r$ profiles in red and black lines. The radiative zone is marked in grey.
	%{Also, I'm puzzled why there is no increase in entropy in model 4 at the shock. Usually it should at least increase first, like in model 10.}
	}
	\label{fig:jupiter}
\end{figure*}

\subsection{Model selection}

To sample a broad range of parameter space, we choose $M_{p}=[0.1,0.3,1,3]$\mj\ to study the accretion of super Neptune, Saturn mass, Jupiter mass and more massive planets. We choose $\dot{M}_{p}=[10^{-3},10^{-2}]$\meyr\ to represent the low and high accretion rates.

Since we do not model the internal structure of the planet (e.g., \citealp{mordasini2012}), which itself depends on the accretion history and shock properties, we cannot give a self-consistent prescription of $r_{\rm in}$ and $p_{\rm in}$ of the forming gas giants. Therefore, we treat $r_{\rm{in}}$ as another parameter and adopt the results that newly formed gas giants with $M_p\gtrsim$\mj\ are expected to have radii of $\sim1.4-4$\rj\ \citep{spiegel2012}. In our simulations, the accretion shock (and the Zel'dovich spike) forms at $r_{p}>r_{\rm{in}}$, where the ram pressure ($\rho v_r^2$) of the infalling gas equals to the pressure of the atmosphere. Note that the free energy available in the shock scales as $M_{p}/r_{p}$ and a larger $r_p$ would largely be equivalent to reducing planet mass (Section \ref{sec:sps} for more discussion). Our choice of $p_{\rm in}$ is also somewhat arbitrary, ranging from $2-10$ bars as $M_{p}$ increases, but we have verified that the results are insensitive to this choice. Using a larger $p_{\rm in}$ will lead to a larger $r_{p}$ as the planetary atmosphere becomes thicker.

We carry out a total of 12 runs, varying $M_{p}$, $\dot{M}_{p}$ and $r_{\rm in}$. The specific run parameters, as well as their major diagnostic properties, are listed in Table \ref{tab:sum}, and they are labeled as models 1 to 12.

\section{Simulation results}\label{sec:result}

In this section, we choose two representative simulations, model 4 and 10, corresponding to accretion onto a Jupiter mass planet with high and low accretion rates. Results from the steady state solutions are shown and analyzed in detail, paying special attention to the post-shock entropy\footnote{More rigorously speaking, we refer to ``post-shock" as the immediate downstream of the Zel'dovich spike.}

\subsection{Overview of simulation results}\label{sec:overview}

Figure \ref{fig:jupiter} shows the steady state solution of models the two models.
In both models, the gas largely free-falls onto the planet as shown in the first panel. The infalling gas is stopped by the planetary atmosphere at $r_{p}$, forming the accretion shock. Gas can be substantially heated to high temperatures at $r_{\rm p}$ by the shock, followed by a rapid fall-off in temperature, known as the Zel'dovich spike. At the Zel'dovich spike, $\tg>\trad$ (see the zoom in plot of the fourth panel), the gas internal energy is quickly converted to the radiation energy, which is then transported outward, seen as the emergence of high radiative energy flux in the second panel from top. The shock is resolved by
8 SMR levels around $r_{p}$ as shown in the third panel. We confirm that our solution is converged in the sense that the post-shock entropy $s_{\rm{ps}}$ (to be defined later) does not differ by much if we use 7 SMR levels. 

From the large to small radii, we can approximately divide our simulation domain into three zones: the pre-shock upstream where gas free falls; the Zel'dovich spike where kinetic energy of the infalling gas is converted to heat and radiation, and 
a radiative zone in the downstream region considered to be a part of the giant planet upper atmosphere.
The three zones are separated by a pre-shock radius $r_{\rm{pre}}$ and a post-shock radius $r_{\rm{ps}}$. We define $r_{\rm{pre}}$ as the radius with the maximum infalling speed and $r_{\rm{ps}}$ as the radius where $T_{\rm{g}}=T_{\rm{rad}}$ in the downstream of the Zel'dovich spike. 

For future convenience, we define radiation mean free path as
\begin{equation}
	\lambda_{\rm{mfp}}=(\kr\rho)^{-1}\ ,
\end{equation}
and we compare $\lambda_{\rm mfp}$ to $r$ in the bottom panel of Figure \ref{fig:jupiter}. The interior of the planet is optically thick with $\lambda_{\rm mfp}/r\ll1$, while the pre-shock region is typically optically thin.\footnote{More rigorously, one may use the $R$ factor (Equation \ref{eqn:R}) to quantify how close the system is to the diffusive or free-streaming limit, and the result is qualitatively the same.} We further define
\begin{eqnarray}
	f_{\rm{r}}&=&\F/(c\E),	\label{eqn:reduced}	\\
	L_{r}&=&4\pi r^{2}\F\ ,
\end{eqnarray}
where $f_{\rm{r}}$ is the reduced radiation flux that characterizes how close radiation transport is to the free-streaming limit, and $L_{r}$ is the outward radiation flux, i.e., luminosity.

Exterior to the planet, the infalling gas undergoes compression and hence adiabatic heating. Additional heating results from the absorption of the outgoing radiation originating from the shock region, maintaining $T_{\rm rad}\sim T_{\rm g}$ \citep{marleau2019}.
When the infalling gas is heated to $T_{\rm{g,dis}}\approx2000$K, \ce{H2} starts to dissociate.
Let us define the number fraction of hydrogen species as,
%\begin{eqnarray}
\begin{equation}
\begin{split}
    \chi_{\ce{H+}}&=\frac{n_{\ce{H+}}m_{\ce{H}}}{\rho X}\ ,	\quad
	\chi_{\ce{H2}}=\frac{2n_{\ce{H2}}m_{\ce{H}}}{\rho X}\ ,	\\
	\chi_{\ce{H}}&=\frac{n_{\ce{H}}m_{\ce{H}}}{\rho X}=1-\chi_{\ce{H+}}-\chi_{\ce{H2}}\ .
\end{split}\label{eqn:chih}
\end{equation}
%\end{eqnarray}
In the third panels, the red and blue lines show the profiles of $\chi_{\ce{H2}}$ and $\chi_{\ce{H+}}$, while $\chi_{\ce{H}}$ can be easily deduced from Equation (\ref{eqn:chih}). The temperatures in these two cases are insufficient to ionize hydrogen. For the high accretion model, \ce{H2} is almost fully dissociated ahead of the shock. For the low accretion model, in contrast, the infalling gas is almost molecular.

It is worth noticing that hydrogen is not ionized at the shock in model 4 but is partially ionized in model 10. This is because the cooling strength in the shock zone is determined by $\kp(a_{\rm{R}}T_{\rm{g}}^{4}-\E)$ and $\kp$ is a strong function of both density and temperature. A high density (from higher accretion rate) at the shock leads to a large $\kp$, thus a relatively low $T_{\rm{g}}$ (at the Zel'dovich spike) is sufficient. Conversely, a low density at the shock would result in a small $\kp$, and hence higher gas temperature that may ionize the hydrogen.
%For more massive gas giants, the Zel'dovich spike region can be hotter and the ionization fraction also becomes higher.

\subsubsection{\ce{H2} dissociation in the pre-shock region}

By comparing the two models, we see that radiation temperature closely follows gas temperature in the pre-shock region in general, except when \ce{H2} gets dissociated. In model 4, the endothermic process increase the gas's heat capacity, leading to $\trad>\tg$ in this region and hence more radiation is absorbed by the infalling gas. This is accompanied by a drop in $L_{r}$ between 2.5-4.7 \rj. In model 10, on the other hand, $L_{r}$ is largely flat in the pre-shock region.

The dissociation of \ce{H2} and absorption of accretion luminosity in the pre-shock has important consequences.
The black line in the fourth panel shows the entropy profile $s$.
When $\tg<T_{\rm{g,dis}}$, the gas radiates away energy as it is compressed, resulting in a slowly decreasing entropy profile as the gas falls inward.
In model 4, upon $\tg$ reaching $T_{\rm{g,dis}}$, the dissociation of \ce{H2} and subsequent energy absorption radiation drives the entropy to rapidly increase from 4.7\rj\ to 2.5\rj\ by almost 10$k_{\rm{B}}/m_{\ce{H}}$. This substantial increase in $s$ is not present in model 10 with only a small fraction of \ce{H2} dissociated.

We note that at the shock, the gas internal energy is quickly converted to the radiation energy, leading to a rapid decrease in entropy.
The post-shock entropy keeps decreasing towards the planet interior accompanied by an outgoing radiation flux. Since there is no radiation flux at the inner boundary, here we name the radiation flux at $r_{\rm{ps}}$ the internal luminosity. The internal luminosity of the planets are marked with black circles in the second panels.

We have also examined simulation results of other models. Generally, results from models 2,3,5,6,11 and 12 are similar to model 4 because a large fraction of \ce{H2} dissociates before the infalling gas hits the planetary atmosphere. They differ in the radii where \ce{H2} dissociation takes place. Results from models 1,7,8 and 9, on the other hand, are similar to model 10, as they all show zero to very low degree of \ce{H2} dissociation in the pre-shock region. They are also very similar to the solutions of perfect gas EoS with $\gamma>4/3$ \citep{marleau2017,marleau2019} as there is no dissociation/ionization except at the shock.

\subsection{Global shock efficiency}

Global shock efficiency quantifies the fractional energy that leaves the system, defined as \citep{marleau2019}
\begin{equation}\label{eqn:eta}
    \eta^{\rm{phy}}=\frac{\dot{E}(r_{\rm{max}})-\dot{E}(r_{\rm{ps}})}{\dot{E}(r_{\rm{max}})},
\end{equation}
where $\dot{E}(r)=-\dot{M}_{p}(\epsilon+\epsilon_{\rm{kin}}+p/\rho+\phi)$ is the total energy flux measured at $r$. A derivation of $\eta^{\rm{phy}}$ can be found in Appendix \ref{appendix:etaphy}.

In long-term planetary evolution, shock efficiency largely determines the initial entropy of the planet and affects the mass-radius relation of the planet.
%Previously, due to a lack of systematic study of the accretion process, 
%the shock efficiency 
Often, it was treated as a free parameter \citep{spiegel2012}, and hence the outcome of the calculations
%making their conclusions 
somewhat hinges on this parameter. Recently, \cite{marleau2017,marleau2019} studied the global shock efficiencies with perfect gas EoS, and conclude that $\eta^{\rm{phy}}\geqslant 97\%$ in their models with $1.3M_{J}\leqslant M_{p}\leqslant10M_{J}$.

\begin{figure*}
	\centering
	\includegraphics[width=\columnwidth]{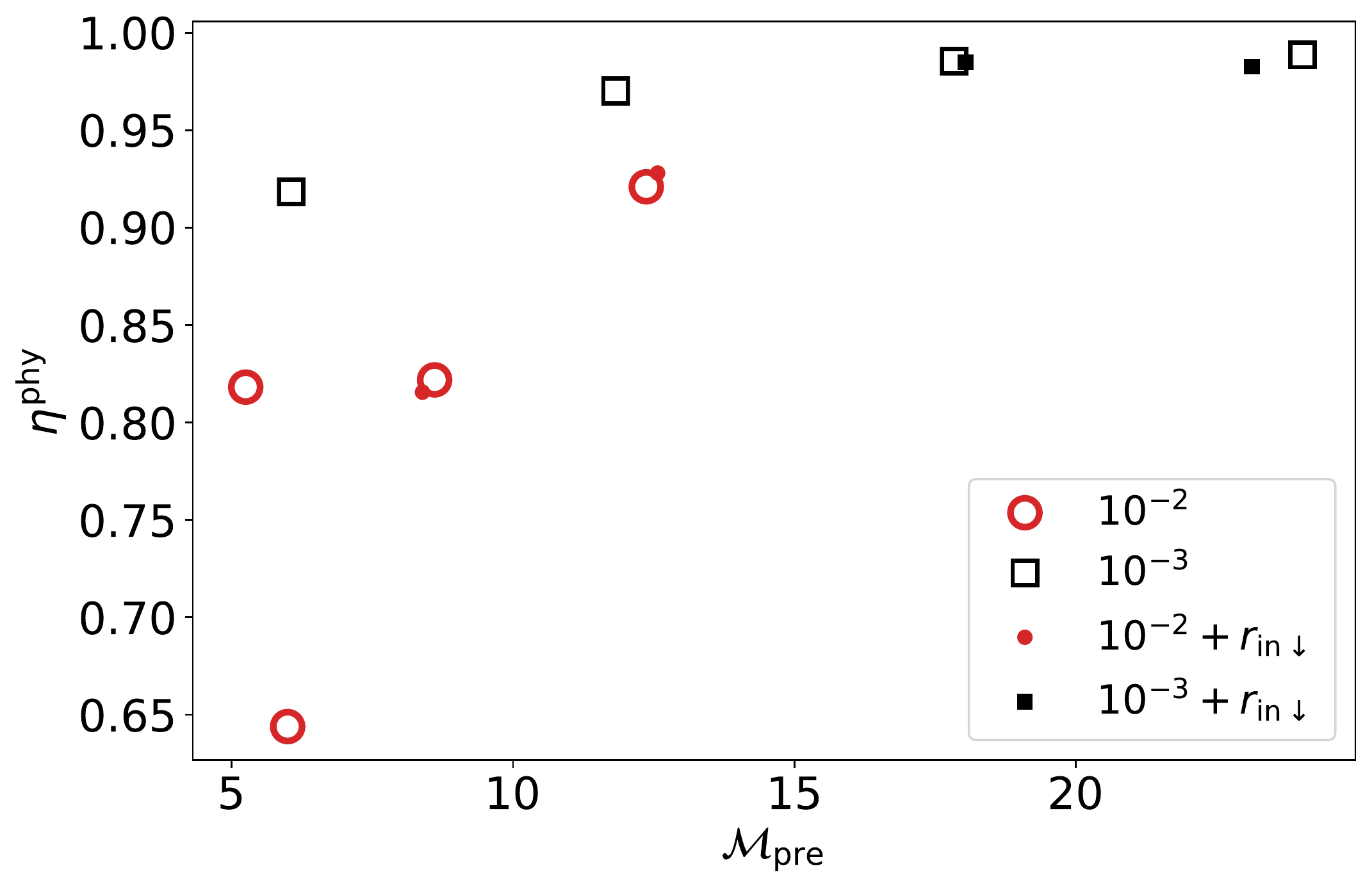}
	\includegraphics[width=\columnwidth]{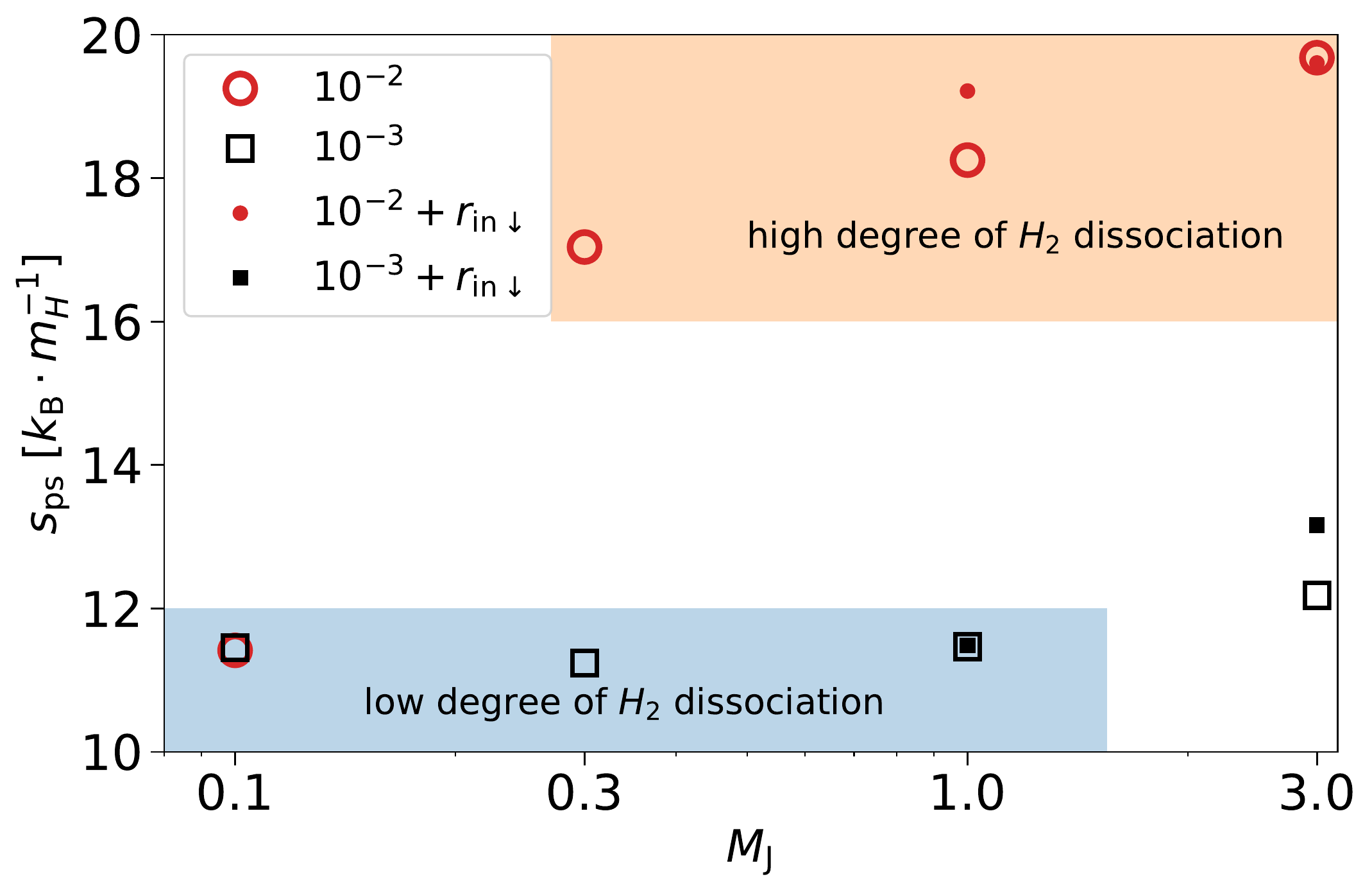}
	\caption{Global shock efficiency (left) and post-shock entropy (right). They are the plots of $\eta^{\rm{phy}}$ v.s. $\mathcal{M}_{\rm{pre}}$, and $s_{\rm{ps}}^{l8}$ v.s. $M_{p}$ in Table \ref{tab:sum}. The red and black colors indicate $10^{-2}$\meyr\ and $10^{-3}$\meyr accretion rate. The filled circles and squares are the results of model 3, 5, 9, and 11 which use smaller $r_{\rm{in}}$.}
	\label{fig:sps}
\end{figure*}

In Figure \ref{fig:sps}, we can see that $64.9\%\leqslant\eta^{\rm{phy}}\leqslant98.92\%$ in our simulations. In particular, models with high accretion rates yield relatively low efficiencies of $64.9\%\leqslant\eta^{\rm{phy}}\leqslant92.81\%$, because the hydrogen dissociation consumes a significant amount of energy and this latent heat is retained in the post-shock region by \ce{H}, lowering $\eta^{\rm{phy}}$. In contrast, the low accretion rate models yield that $91.85\%\leqslant\eta^{\rm{phy}}\leqslant98.92\%,$ as molecular hydrogen is largely retained in the pre-shock region, behaving like perfect gas. Without surprise, $\eta^{\rm{phy}}$ v.s. $\mathcal{M}_{\rm{pre}}$ (the pre-shock gas Mach number) of the low accretion models is largely consistent with Figure 4 of \cite{marleau2019}.

\subsection{Post-shock entropy}\label{sec:sps}

Incorporating all simulation results, we present $s_{\rm ps}$ v.s. $M_{p}$ in Figure \ref{fig:sps}. We find a clear dichotomy, characterized by a high-entropy ``hot group" with $s_{\rm ps}\gtrsim16k_{\rm{B}}/m_{\ce{H}}$ (marked in orange), and a low-entropy ``cold group" with $s_{\rm ps}\lesssim12k_{\rm{B}}/m_{\ce{H}}$ (marked in blue).
By noting the fraction of \ce{H} in the pre-shock and post-shock regions in Table \ref{tab:sum}, we clearly identify that the ``hot group" is associated with models where most \ce{H2} is dissociated before reaching the shock, whereas those without \ce{H2} dissociation before the shock fall into the ``cold group".
We should notice that the post-shock gas will continue to cool in a radiative zone, further decreasing the entropy, and the general zero-age entropy of planets \citep{spiegel2012} should be lower than $s_{\rm{ps}}$.

A few comments are in order concerning the dichotomy.
First, the post-shock entropy generally increases monotonically with increasing accretion rate and planet mass. When choosing intermediate accretion rates for low mass planets $M_p=$0.3-1\mj, one would fill the gap of $s_{\rm ps}$ in Figure \ref{fig:sps} between the two groups. This also applies to model 11 and 12 for massive planets with low accretion rates. On the other hand, the dichotomy remains a valid description, as the ``gap" region corresponds to \ce{H2} being partially dissociated.
Second, \cite{marleau2019} varied $\gamma$ from 1.1 to 1.667 to mimic realistic gas. Their entropy profiles with low $\gamma$ are similar to our results when \ce{H2} dissociates and their high $\gamma$ results are similar to our results when \ce{H2} does not dissociate.
Third, even though $s_{\rm{ps}}$ of massive planets ($>3$\mj) can be low at low accretion rates ($<10^{-3}$\meyr), we should still expect high initial planet entropy because they must build the bulk of their mass through phases of high accretion rates.

Finally, given its importance, we provide an approximate estimate of the condition for which \ce{H2} gets fully dissociated.
This can be obtained by equating radiation temperature from the accretion luminosity at the pre-shock region to $T_{\rm{g,dis}}$.
During dissociation, we have $\F=cf_{r}\E=cf_{r}\ar\trad^{4}$ where $f_{r}\approx1$. Assuming the shock luminosity is much larger than the internal luminosity, we obtain
\begin{eqnarray}
	\eta^{\rm{phy}}\frac{GM_{p}\dot{M_{p}}}{r_{p}}&\approx &4\pi f_{c}r_{p}^{2}cf_{r}\ar T_{\rm{g,dis}}^{4},	\label{eqn:dislimit}\\
	&\approx&6\times10^{-4}f_{c}f_{r}\bigg(\frac{r_{p}}{R_{J}}\bigg)^{2}L_{\odot}.
\end{eqnarray}
We can see that the aforementioned $M_{p}/r_{p}$ is correlated to the dissociation of \ce{H2}. Equation \ref{eqn:dislimit} can be translated to a critical accretion rate for \ce{H2} dissociation, given by
\begin{equation}\label{eqn:critacc}
	\dot{M}_{\rm{crit}}\approx7\times10^{-4}\frac{f_{c}f_{r}}{\eta^{\rm{phy}}}\bigg(\frac{r_{p}}{R_{J}}\bigg)^{3}\bigg(\frac{M_{p}}{M_{J}}\bigg)^{-1}\frac{M_{\oplus}}{\rm yr}.
\end{equation}
We can calculate $\dot{M}_{\rm{crit}}$ by plugging in our assumption of $f_c=1$, and $f_r=1$ with $\eta^{\rm phy}$ from the simulation results. We present $\dot{M}_{p}/\dot{M}_{\rm{crit}}$ in Table \ref{tab:sum}. When $\dot{M}_{p}/\dot{M}_{\rm{crit}}>1$, \ce{H2} is indeed almost fully dissociated; when $\dot{M}_{p}/\dot{M}_{\rm{crit}}\lessapprox1$, \ce{H2} is partially dissociated; when $\dot{M}_{p}/\dot{M}_{\rm{crit}}<0.56$, \ce{H2} largely remains in molecular form in the pre-shock region. In general, we find that this criterion well characterizes the chemical state of the hydrogen in the pre-shock region and separates the ``cold group" and ``hot group" in our simulations (see Figure \ref{fig:sps}).

\section{Summary and discussion}\label{sec:conclusion}

In this letter, we carried out 12 radiation hydrodynamic simulations in 1D with hydrogen EoS to study planetary accretion with a broad range of planet masses and accretion rates. The simulations suggest that hydrogen dissociation in the pre-shocked infalling gas plays an important role in setting the initial condition of giant planets. More specifically, we find

\begin{enumerate}
    \item The post-shock entropy $s_{\rm ps}$ of the 12 simulations generally fall into two groups (Figure \ref{fig:sps}): a ``cold" group with low degree of \ce{H2} dissociation and a ``hot" group with high degree of \ce{H2} dissociation.
    \item The global shock efficiency $\eta^{\rm{phy}}$ - the fraction of the accretion energy that is radiated away - can be lowered by hydrogen dissociation.
    \item There is a critical accretion rate $\dot{M}_{\rm{crit}}$ above which \ce{H2} gets largely dissociated in the infalling gas, which is given by Equation \ref{eqn:critacc}.
\end{enumerate}

We note that the post-shock entropy (Figure \ref{fig:sps}) should be higher than the entropy of zero-age planets because the newly accreted gas can still radiate energy away. 
In addition, our simulations represent individual snapshots of planetary accretion shocks, while the final result should depend on the accretion history, especially over the period when giant planets build most of their masses.
Our results of the global shock efficiencies and post-shock entropy can serve as more realistic inputs for more detailed modeling of the structure and evolution of accreting planets \citep{mordasini2012,berardo2017}, which eventually determines the initial conditions of gas giants.
%As mentioned in Section \ref{sec:sps}, we expect high initial planet entropy for massive planet ($>$3\mj) and it should be verified in the future with a more thorough model.

As a first approximation, we have assumed that accretion is spherically symmetric. While we expect the actual accretion process could be more complex through a CPD \citep{takasao2021}, we note that existing studies of CPDs have yet to self-consistently incorporate the major physical ingredients including radiation with realistic EoS and magnetic fields. Despite the major uncertainties in our ignorance of how gas giants accrete, our results also imply that whether the hydrogen is accreted in molecular or atomic form likely has significant impact in setting the initial condition of gas giants.

Current imaging surveys may be approaching the limit to detect more young gas giants in PPDs, especially in systems with large inner cavities (\citealp{asensio2021}, but see \citealp{sanchis2020}). There are also giant planets with predicted dynamical mass awaiting for direct imaging detection (e.g., \citealp{dong2015,maire2017,brown2021}), which are ideal tools to test the evolution models of gas giants. The discovery space will be greatly enhanced with the upcoming James Webb Space Telescope (JWST, \citealp{carter2021}), the Extremely Large Telescope (ELT, \citealp{Carlomagno2020}), and the Chinese Space Station Telescope (CSST) which, together with modeling effort, will likely yield a more decisive picture of gas giant formation.

\begin{acknowledgments}
We are indebted to Gabriel-Dominique Marleau for valuable comments on a preliminary version of the paper, and thank Yuhiko Aoyama, Ruobing Dong, Chris Ormel, and Shude Mao for helpful discussions. We also thank the anonymous referee whose suggestions improved the quality of this Letter. This work is supported by the National Key R\&D Program of China (No.2019YFA0405100). ZC is grateful to the CITA National Postdoctoral Fellowship, Tsinghua Astrophysics Outstanding Fellowship, and Shuimu Tsinghua Scholar Program. 
\end{acknowledgments}

\software{{\tt Guangqi} \citep{chen2021}, {\tt Matplotlib} \citep{hunter2007}, {\tt Petsc} \citep{petsc-efficient,petsc-user-ref}}


\bibliography{acc_planet}
\bibliographystyle{aasjournal}

\begin{appendix}
\section{Equation of state}\label{appendix:eos}

The general EoS can be derived based on the partition functions of the underlying species. The partition functions of all the species used in this work are given by
\begin{eqnarray}
	Z_{\ce{H2}}&=&\left(\frac{2\pi m_{\ce{H2}}k_{\rm{B}}\tg}{h^2}\right)^{3/2},	\\
	Z_{\ce{H}}&=&\left(\frac{2\pi m_{\ce{H}}k_{\rm{B}}\tg}{h^2}\right)^{3/2}\exp\left(-\frac{\phi_{\rm{dis}}}{2k_{\rm{B}}\tg}\right),	\\
	Z_{\ce{H+}}&=&\left(\frac{2\pi m_{\ce{H+}}k_{\rm{B}}\tg}{h^2}\right)^{3/2}\exp\left(-\frac{\phi_{\rm{dis}}+2\phi_{\rm{ion}}}{2k_{\rm{B}}\tg}\right),	\\
	Z_{\ce{e-}}&=&\left(\frac{2\pi m_{\ce{e-}}k_{\rm{B}}\tg}{h^2}\right)^{3/2},	\\
	Z_{\ce{He}}&=&\left(\frac{2\pi m_{\ce{He}}k_{\rm{B}}\tg}{h^2}\right)^{3/2},
\end{eqnarray}
where $m_{\ce{H2}}$, $m_{\ce{H}}$, $m_{\ce{H+}}$, $m_{\ce{e-}}$, and $m_{\ce{He}}$ are the masses of the elements. $\phi_{\rm{dis}}=7.17\times10^{-12}$ erg and $\phi_{\rm{ion}}=2.18\times10^{-11}$ erg are the binding energy of \ce{H2} and \ce{H}.

In the above, we do not consider the spin and atomic structure of any species, and have ignored the rotational and vibrational degrees of freedom of \ce{H2}. Incorporating these degrees of freedom would lead to much complex partition functions, yet because the energy needed to excite these transitions are small compared to the binding energy, we expect the resulting impact to be relatively minor. Moreover, we need to use a tabulated EoS if we were to include them to the partition function, which would make our calculations inefficient and less accurate.

In many of the simulations, \ce{H2} reforms in the post-shock region. We note that the underlying assumption with our treatment of the general EoS is chemical equilibrium, implying that \ce{H2} formation is instantaneous based on the Saha equation. In reality, given that the post-shock gas density is high ($\rho_{\rm{ps}}>10^{-8}$ \gcmc), we anticipate that efficient \ce{H2} formation can be achieved through the three-body \ce{H2} formation channel on a timescale of less than one minute \citep{omukai2005}, which is much shorter than the free-fall time.

\section{Global shock efficiency}\label{appendix:etaphy}

Global shock efficiency $\eta^{\rm{phy}}$ quantifies the fraction of the energy that leave the system, in our case the radiation energy. For steady state solutions, Equation \ref{eqn:energyhydro} and \ref{eqn:energyrad} become,

\begin{eqnarray}
	\frac{1}{r^{2}}\pdv{r}[r^{2}(\epsilon_{\rm kin}+\epsilon+p)v]&=&\mathbb{G}-\rho v\nabla\phi,	\label{eqn:steady1}\\
	\frac{1}{r^{2}}\pdv{r}(r^{2}\F)&=&-\mathbb{G},	\label{eqn:steady2}
\end{eqnarray}
where $\phi=-GM_{p}/r$ is the gravitational potential energy. We can multiply by $r^2$ and add up the two equations above, recognizing that $\dot{M}_{p}=-4\pi\rho r^{2}v$ is constant in steady state and $4\pi r^{2}\F=L_{r}$ is the luminosity, to obtain
\begin{equation}\label{eqn:etotal}
	\dot{M}_{p}\pdv{r}(\epsilon+\epsilon_{\rm{kin}}+p/\rho+\phi)+\pdv{L_{r}}{r}=0\ .
\end{equation}
Equation \ref{eqn:etotal} illustrates the balance of the radiative flux and the total energy flux $\dot{E}(r)=-\dot{M}_{p}(\epsilon+\epsilon_{\rm{kin}}+p/\rho+\phi)$ in steady state. Since radiation energy is the one that leaves the system, 
the fraction of accretion energy radiated away, which is defined as the shock efficiency $\eta^{\rm{phy}}$, is given by
\begin{equation}\label{eqn:etaphy}
	\eta^{\rm{phy}}=\frac{L_{\rm{r}}(r_{\rm{max}})-L_{\rm{r}}(r_{\rm{ps}})}{\dot{E}(r_{\rm{max}})}=\frac{\dot{E}(r_{\rm{max}})-\dot{E}(r_{\rm{ps}})}{\dot{E}(r_{\rm{max}})}\le1\ .
\end{equation}
Consequently, $1-\eta^{\rm{phy}}$ is the fraction of the energy retained after the shock.

\end{appendix}


\end{CJK*}
\end{document}